%% file: condmat.tex
\documentstyle[aps,multicol,epsfig]{revtex}

\begin{document}

\draft
 
\title{Universal free energy correction for the two-dimensional one-component plasma}

\author{Bernard Jancovici\footnote{Electronic Address: Bernard.Jancovici@th.u-psud.fr} 
and Emmanuel Trizac
\footnote{Electronic Address: Emmanuel.Trizac@th.u-psud.fr}}

\address{
Laboratoire de Physique 
Th\'eorique\footnote{Unit\'e Mixte de Recherche UMR 8627 du CNRS}, 
B\^atiment 210\\ Universit\'e de Paris-Sud, 91405 Orsay Cedex, France
}
 
\date{\today}

\maketitle
\begin{abstract}
The universal finite-size correction to the free energy of a 
two-dimensional Coulomb 
system is checked in the special case of a one-component plasma
on a sphere. The correction is related to the known second moment of the short-range
part of the direct correlation function for a planar system. 
\end{abstract}




\section{Introduction}

Two-dimensional Coulomb systems are models which have attracted some
attention. On a two-dimensional manifold, a Coulomb system is a system
made of particles interacting through the corresponding Coulomb
potential plus perhaps some short-range interaction. In a plane, the
Coulomb interaction energy of two particles of charges $q$ and $q'$,
separated by a distance $r$, is defined as $-qq'\ln (r/L)$, where $L$ is   
some (irrelevant) length. 

Some time ago, it has been shown that the free energy of such systems
has a universal finite-size correction \cite{JancoManiPisa}
 very similar (except for its
sign) to the one which occurs in a system with short-range forces at a
critical point \cite{Cardy}: for a finite Coulomb system of characteristic size
$R$, the free energy $F$ has the large-$R$ behaviour 

\begin{equation}
\beta F = AR^2 + BR + \frac{\chi}{6} \,\ln R + \cdots, 
\label{eq:finitesize}
\end{equation}
where $\beta$ is the inverse temperature. $A$ and $B$ are
non-universal constants describing the bulk and boundary contributions,
respectively. $(\chi/6)\ln R$ is the universal correction, depending
only on the Euler number $\chi$ which describes the topology of the
manifold on which the system lives. However, the general derivation 
\cite{JancoManiPisa} of (\ref{eq:finitesize}) had some heuristic features. 
The purpose of the present paper is
to check (\ref{eq:finitesize}) in the special case of a one-component plasma on the
surface of a sphere, by a different method.

The one-component plasma is a system made of one species of
point-particles of charge $q$ in a uniform neutralizing background. On a
sphere of radius $R$, the interaction between two particles can be
chosen \cite{Caillol,CaillolLevesque}
as $-q^2\ln [(2R/L)\sin(\psi/2)]$, where $\psi$ is the
angular distance (seen from the sphere centre) between the two
particles. There are also particle-background and background-background
interactions. A dimensionless coupling constant is $\Gamma=\beta q^2$.

A sphere has no boundaries and its Euler number is $\chi=2$.
Furthermore, for a given particle density, $R^2$ is proportional 
to the number of particles $N$. Thus, expansion (\ref{eq:finitesize}) 
becomes 
\begin{equation}
\beta F = CN + \frac{1}{6}\,\ln N + \cdots, 
\label{eq:Ffinitesize}
\end{equation}
where $C$ is a constant. The model is exactly solvable \cite{Caillol} when
$\Gamma=2$ and it can be checked \cite{JancoManiPisa} that (\ref{eq:Ffinitesize}) 
is obeyed in that
case. Also, exact calculations \cite{Tellez} 
for finite values of $N$ at
$\Gamma=4$ and $\Gamma=6$ are well fitted by (\ref{eq:Ffinitesize}).  

The present derivation of (\ref{eq:Ffinitesize})
 relies on a recent result \cite{Kalinay} about the
direct correlation function $c(r)$ of the plane one-component plasma. 
By a diagrammatic analysis, it has been shown in ref. \cite{Kalinay} 
that the second
moment of the short-range part $c_{_{SR}}(r)$ has the simple value
\begin{equation}
n^2\int c_{_{SR}}(r)\,r^2\, d^2{\bf r}= \frac{1}{12\pi}. 
\label{eq:sumrule}
\end{equation}
A remarkable feature of (\ref{eq:sumrule})
is its universality, in the sense
that it is independent of the coupling constant $\Gamma$. It will now be
shown how (\ref{eq:sumrule}) leads to (\ref{eq:Ffinitesize}). 
More specifically, we show how (\ref{eq:sumrule}) implies the
finite-size correction to the chemical potential $\mu = \partial F/\partial N$:
\begin{equation}
\beta \mu = \beta \mu_{\infty} \,+\,\frac{1}{6N}\,+\,\ldots
\label{eq:mufinitesize}
\end{equation}
This derivation bears some similarity with another
one about the two-component plasma \cite{Janco,JancoKali}.

\section{Density functional theory approach}

We consider the OCP of average density $n_s$ on the sphere of radius $R$
(with a corresponding number of particles $N=4 \pi R^2\, n_s$).
Introducing the stereographic projection of the sphere onto the plane 
${\cal P}$ tangent to its south pole (see figure 1), we map the 
homogeneous OCP on the sphere
onto a modified inhomogeneous plasma on the plane, with local particle density
\begin{equation}
n({\bf r}) = n_s \, \left( 1+ \frac{r^2}{4 R^2}\right)^{-2}.
\label{eq:density}
\end{equation}
In terms of planar coordinates ${\bf r}_1$ and ${\bf r}_2$, the interaction potential
between two particles on the sphere with angular distance $\psi_{12}$ can be written
as the sum of the planar two dimensional Coulomb potential
$v_p({\bf r},{\bf r'}) = -q^2 \ln[|{\bf r}-{\bf r'}|/L]$ and one-body terms since:
\begin{equation}
-\ln\left[ \frac{2R}{L} \,\sin\left( \frac{\psi_{12}}{2}
\right)\right] = -\ln\left[\frac{|{\bf r}_1-{\bf r}_2|}{L} \right] \,+\,
\frac{1}{2}\,\ln\left(1+\frac{r_1^2}{4 R^2}   \right) \,+\,
\frac{1}{2}\,\ln\left(1+\frac{r_2^2}{4 R^2}   \right).
\label{eq:potential}
\end{equation}
The two one-body terms appearing on the right hand side of eq. (\ref{eq:potential}),
as well as the metric, create a central potential and
we can consider the projected planar system as a 
OCP interacting through
the standard pair potential $v_p$, in an external one-body central
potential $V_R^N(r)$. The latter acts as a confining mechanism 
ensuring the proper density given by eq. 
(\ref{eq:density}), and its detailed form need not be precised.
Without background, the free energy ${\cal F}'$ 
for the set of particles with pair potential $v_p$ can be formally
expanded in a Mayer diagrammatic representation \cite{Kalinay}, with a leading 
term $(1/2) \int n({\bf r})  v_p({\bf r},{\bf r}') n({\bf r}')
d^2{\bf r}\, d^2{\bf r'}$ for the excess (over ideal) part of ${\cal F}'$.
The presence of a neutralizing background cancels this mean-field
electrostatic term and the intrinsic free energy functional 
of the inhomogeneous OCP becomes:
\begin{equation}
{\cal F}[n({\bf r})] = {\cal F}'[n({\bf r})] - \frac{1}{2} \int n({\bf r})\, 
v_p({\bf r},{\bf r}')\, n({\bf r}') \, d^2{\bf r}\, d^2{\bf r'}.
\label{eq:free}
\end{equation}

The local chemical potential reads 
\begin{equation}
\mu({\bf r})  = \frac{\delta {\cal F}[n]}{\delta n({\bf r})} 
     = \frac{\delta {\cal F}'[n]}{\delta n({\bf r})} - \int v_p({\bf r},{\bf r}')\,
     n({\bf r}')\, d^2{\bf r}' 
\label{eq:chemical}
\end{equation}
and the second functional derivative of ${\cal F}$ yields:
\begin{eqnarray}
\beta\frac{\delta \mu({\bf r})}{\delta n({\bf r}')} &=& 
\frac{\delta^2 {\cal \beta F}'[n]}{\delta n({\bf r})\delta n({\bf r}')}
- \beta v_p({\bf r},{\bf r'}) \\
&=&  -c({\bf r},{\bf r'}) + \frac{\delta({\bf r}-{\bf r}')}{n({\bf r})}
- \beta v_p({\bf r},{\bf r'})
\\
&=& -c_{_{SR}}({\bf r},{\bf r}') + \frac{\delta({\bf r}-{\bf r}')}{n({\bf r})},
\label{eq:secondder}
\end{eqnarray}
where the variations of the excess contribution to ${\cal F}'$ give rise
to the usual direct correlation function \cite{Hansen}, having
a short-range part given by 
\begin{equation}
c_{_{SR}}({\bf r},{\bf r}') = c({\bf r},{\bf r}')\,+\,\beta v_p({\bf r},{\bf r}').
\end{equation}
Note that
the chemical potential of the OCP on the sphere coincides with $\mu(0)$
for the optimum density profile (\ref{eq:density}).

Equation (\ref{eq:secondder}) emphasizes the short range dependence of the
chemical potential on a density perturbation. Consequently,
$\mu(0)$ is the same in a finite $N$-particle OCP
in the central potential $V_R^N(r)$ and in the limit $N\to \infty$ 
with an external potential $V_R^\infty(r)$ ensuring the same density variation
around the origin as expression (\ref{eq:density}), namely:
\begin{equation}
n({\bf r}) = n_s \left(1-\frac{r^2}{2 R^2}\right) \,+\, \ldots
\end{equation}
For the purpose of the present analysis, it is sufficient to truncate
(\ref{eq:density}) after second order in $r$, as becomes clear below.
The knowledge of the finite-size correction to the chemical potential 
for the OCP on the sphere then amounts to computing the shift $\delta \mu(0)$
induced by switching $V_R^\infty(r)$ starting from the infinite homogeneous planar
OCP with density $n_s$ (corresponding to the stereographic projection 
of the ``spherical'' plasma in the thermodynamic limit $R \to \infty$).
The density variation caused by the addition of $V_R^\infty(r)$ reads
$\delta n({\bf r}) \simeq -n_s r^2/(2 R^2)$ and induces the shift
\begin{eqnarray}
\beta\,\delta\mu(0) &=& \int \left[ 
-c_{_{SR}}({\bf r}) + \frac{\delta({\bf r})}{n({\bf r})} \right] \,
\delta n({\bf r}) \, d^2{\bf r}\\
&=& \frac{n_s}{2 R^2}\,\int c_{_{SR}}(r)\, r^2 \, d^2{\bf r},
\end{eqnarray}
where the direct correlation function to be considered is that of the homogeneous
reference planar OCP. From the sum rule (\ref{eq:sumrule}), we finally obtain:
\begin{equation}
\beta\,\delta\mu(0) \,=\, \frac{1}{24 \pi n_s R^2} \,=\, \frac{1}{6N},
\end{equation}
and eq. (\ref{eq:mufinitesize}) is recovered.


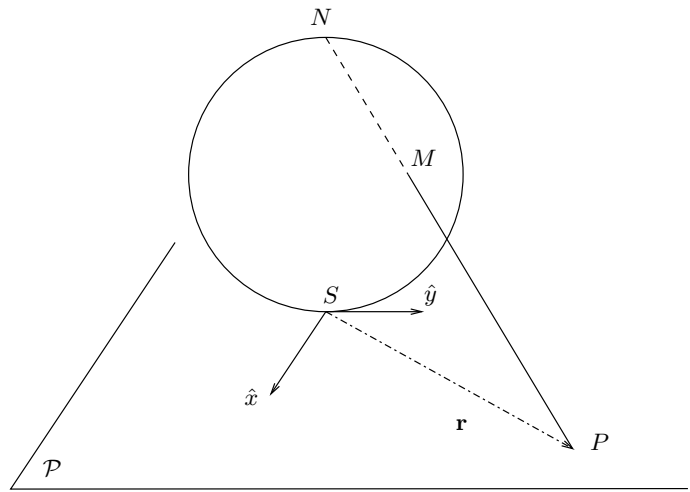
\begin{figure}
$$\input{stereopetit.pstex_t}$$
\vskip 1cm
\caption{ Stereographic projection from the North pole onto
the plane ${\cal P}$. A point $M$ on the sphere is projected onto
$P$, with Cartesian coordinates ${\bf r}$ (${\bf r} =\bbox{0}$ at the South pole
$S$).
}
\end{figure}


\end{document}

%% file: stereopetit.pstex_t
\begin{picture}(0,0)%
\epsfig{file=stereopetit.pstex}%
\end{picture}%
\setlength{\unitlength}{3947sp}%
\begingroup\makeatletter\ifx\SetFigFont\undefined%
\gdef\SetFigFont#1#2#3#4#5{%
  \reset@font\fontsize{#1}{#2pt}%
  \fontfamily{#3}\fontseries{#4}\fontshape{#5}%
  \selectfont}%
\fi\endgroup%
\begin{picture}(4278,3054)(4495,-6515)
\put(6400,-3572){\makebox(0,0)[lb]{\smash{\SetFigFont{9}{10.8}{\rmdefault}{\mddefault}{\updefault}\special{ps: gsave 0 0 0 setrgbcolor}$N$\special{ps: grestore}}}}
\put(7116,-5304){\makebox(0,0)[lb]{\smash{\SetFigFont{9}{10.8}{\rmdefault}{\mddefault}{\updefault}\special{ps: gsave 0 0 0 setrgbcolor}$\hat y$\special{ps: grestore}}}}
\put(7029,-4459){\makebox(0,0)[lb]{\smash{\SetFigFont{9}{10.8}{\rmdefault}{\mddefault}{\updefault}\special{ps: gsave 0 0 0 setrgbcolor}$M$\special{ps: grestore}}}}
\put(8157,-6242){\makebox(0,0)[lb]{\smash{\SetFigFont{9}{10.8}{\rmdefault}{\mddefault}{\updefault}\special{ps: gsave 0 0 0 setrgbcolor}$P$\special{ps: grestore}}}}
\put(4720,-6406){\makebox(0,0)[lb]{\smash{\SetFigFont{9}{10.8}{\rmdefault}{\mddefault}{\updefault}\special{ps: gsave 0 0 0 setrgbcolor}${\cal P}$\special{ps: grestore}}}}
\put(6479,-5321){\makebox(0,0)[lb]{\smash{\SetFigFont{9}{10.8}{\rmdefault}{\mddefault}{\updefault}\special{ps: gsave 0 0 0 setrgbcolor}$S$\special{ps: grestore}}}}
\put(5987,-5959){\makebox(0,0)[lb]{\smash{\SetFigFont{9}{10.8}{\rmdefault}{\mddefault}{\updefault}\special{ps: gsave 0 0 0 setrgbcolor}$\hat x$\special{ps: grestore}}}}
\put(7314,-6120){\makebox(0,0)[lb]{\smash{\SetFigFont{9}{10.8}{\rmdefault}{\mddefault}{\updefault}\special{ps: gsave 0 0 0 setrgbcolor}${\bf r}$\special{ps: grestore}}}}
\end{picture}